\documentclass[preprint]{revtex4}
\usepackage{amsfonts}
\usepackage{appendix}
\usepackage{amssymb}
\usepackage{amsmath}
\usepackage{graphicx}

\begin{document}

\newcommand{\beq}{\begin{equation}}
\newcommand{\eeq}{\end{equation}}
\newcommand{\ket}[1]{\left\vert #1 \right\rangle}
\newcommand{\bra}[1]{\left\langle #1 \right\vert}
\newcommand{\bea}{\begin{eqnarray}}
\newcommand{\eea}{\end{eqnarray}}
\newcommand{\id}{\mathbb{I}}

\title{Effective and efficient resonant transitions in periodically
modulated quantum systems}
\author{Isabel Sainz$^1$ , Andr\'es Garc\'ia$^2$  and Andrei B. Klimov$^1$  }
\date{\today}

\begin{abstract}
We analyse periodically modulated quantum systems with $SU(2)$ and $SU(1,1)$
symmetries. Transforming the Hamiltonian into the Floquet representation we
apply the Lie transformation method, which allows us to classify all
effective resonant transitions emerging in time-dependent systems. In the
case of a single periodically perturbed system, we propose an explicit
iterative procedure for the determination of the effective interaction
constants corresponding to every resonance both for weak and strong
modulation. For coupled quantum systems we determine the efficient resonant
transitions appearing as a result of time modulation and intrinsic
non-linearities.
\end{abstract}

\address{$^1$  Departamento de F\'sica, Universidad de Guadalajara, Revoluci\'on 1500, 44420, Guadalajara, Jal., M\'exico\\
$^2$ Departamento de Matem\'aticas, Universidad de Guadalajara, Revoluci\'on 1500, 44420, Guadalajara, Jal., M\'exico}

\maketitle

\section{Introduction}

Effective transitions in time-dependent quantum systems have been
extensively studied since the classical paper \cite{autler}, later
generalized in \cite{shirley} and widely applied for the description of
atomic dynamics in external fields \cite{yabuzaki}-\cite{cr}; and in more
involved periodically perturbed quantum systems \cite{SQmicrowave}-\cite%
{dodonov}. Effective transitions are described by operators that: i) become
time-independent (resonant) in an appropriate reference frame under certain
relations between the system%
\'{}%
s frequencies (resonant conditions); ii) are not present in the original
Hamiltonian; iii) disappear in the Rotating Wave Approximation (RWA). Such
resonant terms (later referred to as \textit{resonances}) naturally appear
in the effective Hamiltonian in the weak interaction limit where the
counter-rotating (CR) contributions, rapidly oscillating in \textit{any}
reference frame terms, are perturbatively taken into account. The most
famous examples of time-dependent systems with an infinite number of
effective resonances are the Rabi model in classical field \cite{autler},
\cite{shirley} and the parametric quantum oscillator \cite{OscPar}. Even in
these simplest systems, where the Hamiltonian is a linear form on the $su(2)$
and $su(1,1)$ Lie algebras,\ it turns out that the general expressions for
the effective interaction constants and the frequency shifts in the vicinity
of each resonance, are not easy to obtain.

The situation becomes even more complicated when two coupled quantum systems
are subjected to time-dependent periodic perturbations e.g., as in the
quantum Rabi model with modulated coupling and/ or frequency. In these types
of models, CR terms (in the absence of external fields) are responsible for
several physical effects such as: multiphoton atom-field interactions in the
Rabi model \cite{resonance,3fot}, an improvement of a qubit photodetector
readout \cite{photodetector}, the excitation of several atoms by a single
photon \cite{1f2a,resonance}, and several other effective processes now
experimentally achievable in solid state circuit QED setups \cite%
{exp1,reviewExp,circuitQED}. An additional periodic excitation makes the
situation even richer, leading to e.g., the enhancement of CR interactions
in the Rabi model \cite{cr}, the generation of specific non-classical photon
states \cite{SQmicrowave}, the emergence of non-linear spin-boson couplings
\cite{MC}, the appearance of quantum to classical phase-transitions \cite%
{qtoc}, lasing with a single atom \cite{1atomLaser}, simulation of the
anisotropic Rabi model \cite{AnisotropicRabi}, or the dynamical Casimir
effect \cite{casimir}.

One of the possibilities to construct a perturbation theory that unveils the
effective resonant interactions, is the Lie transformation method \cite%
{smallrot}. Such a method consists in order-by-order elimination of CR terms
by a specific set of transformations, which particular form directly follows
from the algebraic structure of the original Hamiltonian \cite{resonance}.
The advantage of this approach consists in a rather simple and systematic
procedure for obtaining the general form of effective resonance terms and
the order of the corresponding effective coupling constants.

The aim of the present paper is to provide a systematic approach to the
analysis of effective resonant transitions in quantum systems obeying the $%
SU(2)$, $SU(1,1)$ and $H(1)$ symmetries with periodically modulated
frequencies and/or coupling constants. We construct the Lie-type all-order
perturbation theory allowing to determine the order of every possible
resonance that may emerge in the effective Hamiltonians. We consider both
single and coupled quantum systems and determine the efficient resonant
transitions emerging as a combination of the time modulation and intrinsic
non-linearities, especially relevant in interacting systems.

In Sec.II we outline the Lie transformation method in an example of a single
periodically perturbed $su(2)$/$su(1,1)$ system and provide not only the
order of the effective resonance terms \cite{pla} but the explicit iterative
procedure for determining the effective interaction constants both for the
weak and strong modulation. In Sec.III we analyze coupled time-dependent
quantum systems and discuss types of \textit{efficient} resonances proper to
different symmetries of the interacting systems.

\section{Single periodically modulated quantum system}

\label{1sys}

\subsection{General settings}

Let us consider a quantum system described by the following time-dependent
Hamiltonian (the case where only the frequency of the system is modulated is
considered in Sec. 2.4),
\begin{equation}
H_{\pm }(t)=\omega X_{0}+2g_{0}\cos \left( \nu t\right) X_{0}+2g_{1}\cos
\left( \nu t\right) \left( X_{+}+X_{-}\right) ,\;\omega ,\nu >0,
\label{H1dgen}
\end{equation}%
where the operators $X_{\pm },X_{0}$, satisfy the following commutation
relations%
\begin{equation}
\left[ X_{0},X_{\pm }\right] =\pm X_{\pm },\,\quad \left[ X_{+},X_{-}\right]
=\pm 2X_{0},  \label{CR}
\end{equation}%
the signs $\pm $ correspond to the $su(2)$ and $su(1,1)$ algebras
respectively. In the interaction picture the Hamiltonian takes the form%
\begin{eqnarray}
H_{int\pm }(t) &=&g_{0}\left( e^{i\nu t}+e^{-i\nu t}\right) X_{0}  \label{Xt}
\\
&&+g_{1}\left( e^{i(\omega +\nu )t}X_{+}+e^{i(\omega -\nu
)t}X_{+}+h.c.\right) ,  \label{Hint1d0}
\end{eqnarray}%
where the terms $\sim e^{i(\omega -\nu )t}X_{+}+h.c.$ become time
independent if $\omega =\nu $ and correspond to the principal resonance,
while the counter-rotating (CR) terms $\sim \cos \left( \nu t\right) X_{0}$
and $e^{i(\omega +\nu )t}X_{+}+h.c.$ rapidly oscillate for any relation
between the frequencies.

The CR terms are neglected in the zero-order approximation \ when $g_{1}\sim
g_{0}\ll \omega ,\nu $ (RWA) and the Hamiltonian (\ref{Xt})-(\ref{Hint1d0})
acquires a simple form
\begin{equation*}
H_{int\pm }\approx g_{1}\left( e^{i(\omega -\nu )t}X_{+}+h.c.\right) .
\end{equation*}%
In the opposite limit, $g_{1}\ll g_{0}\sim \omega ,\nu $ the zero-order
approximation gives the diagonal Hamiltonian%
\begin{equation*}
H_{int\pm }\approx g_{0}\left( e^{i\nu t}+e^{-i\nu t}\right) X_{0},
\end{equation*}%
characterized by a trivial dynamics. The CR terms in (\ref{Xt})-(\ref%
{Hint1d0}) lead to the emergence of non-trivial resonant transitions, not
explicitly present in the original Hamiltonian. It is well known that in the
case of $H(1)$ symmetry, $X_{+}=a^{\dag },~X_{-}=a$, $X_{0}=a^{\dag }a$, no
resonances additional to the principal one $\sim ae^{i(\omega -\nu )t}$
appear.

In order to develop the Lie-perturbation theory, that allows to describe all
possible effective resonances both in the limits $g_{0,1}\ll \omega ,\nu $
and $g_{1}\ll g_{0}\sim \omega ,\nu $, we introduce the Euclidean algebra
operators $E,E^{\dag },E_{0}$, obeying the commutation relations
\begin{equation}
\lbrack E_{0},E]=-E,~[E_{0},~E^{\dag }]=E^{\dag },~[E^{\dag },E]=0,
\label{E}
\end{equation}%
where $E^{\dag }E=EE^{\dag }=I$. Then, the Hamiltonian (\ref{H1dgen}) can be
put in a one-to-one correspondence with the following time-independent
(Floquet) form \cite{pla}%
\begin{eqnarray}
H_{\pm }^{F} &=&\omega X_{0}+\nu E_{0}+g_{0}\left( E+E^{\dag }\right) X_{0}
\notag \\
&&+g_{1}\left( E+E^{\dag }\right) \left( X_{+}+X_{-}\right) ,  \label{H1dop0}
\end{eqnarray}%
where the Euclidean operators $(E,E^{\dag },E_{0})$ describe a
\textquotedblleft classical\textquotedblright\ field interacting with the $X$%
-system. The time-dependent Hamiltonian (\ref{H1dgen}) is recovered from (%
\ref{Hint1d0}) by transforming it to the frame rotating with the
\textquotedblleft classical frequency\textquotedblright\ $\nu $,
\begin{eqnarray}
H_{int\pm }^{F} &=&g_{0}\left( e^{i\nu t}E^{\dag }+e^{-i\nu t}E\right) X_{0}
\notag \\
&&+g_{1}\left( e^{it\left( \omega +\nu \right) }E^{\dag }X_{+}+e^{it\left(
\omega -\nu \right) }EX_{+}+h.c.\right) ,  \notag
\end{eqnarray}%
with a subsequent averaging over the eigenstates of $E,E^{\dag }$ operators
(phase-states), $E\left\vert \phi \right\rangle =e^{-i\phi }\left\vert \phi
\right\rangle $,
\begin{equation}
\left\vert \phi \right\rangle =\lim_{N\rightarrow \infty }\frac{1}{\sqrt{2N+1%
}}\sum_{n=-N}^{N}e^{-in\phi }\left\vert n\right\rangle ,  \label{phaseSt}
\end{equation}%
where $E_{0}\left\vert n\right\rangle =n\left\vert n\right\rangle $, and
setting the initial phase $\phi =0$ without loss of generality. The CR terms
are now identified with $\sim \left( E+E^{\dag }\right) X_{0}$ and $\sim
\left( EX_{-}+E^{\dag }X_{+}\right) $.

\subsection{Effective Hamiltonian, $g_{0,1}\ll \protect\omega ,\protect\nu $}

We start with the most complicated limit $g_{0}\sim g_{1}\ll \omega ,\nu $,
when the contributions of diagonal and non-diagonal CR terms in (\ref{Xt})-(%
\ref{Hint1d0}) are of the same order. The CR terms appearing in the Floquet
Hamiltonian (\ref{H1dop0}) can be removed order - by - order by applying a
set of Lie-type transformations according to the general scheme \cite%
{resonance}, \cite{smallrot} as shown in Appendix A.

The common feature of all of these transformations (small rotations) is
their form
\begin{equation}
V_{k}=e^{\varepsilon _{k}\left( Z_{+}^{(k)}-Z_{-}^{(k)}\right) },  \label{V}
\end{equation}%
where $\varepsilon _{k}\ll 1$ are some appropriate small parameters, under
the condition that the Hamiltonian, which is transformed by (\ref{V}),
should contain the term $Z_{0}^{(k)}$ such that $\left[ Z_{0}^{(k)},Z_{\pm
}^{(k)}\right] =\pm Z_{\pm }^{(k)}$. The resonance expansion is obtained as
a power series of the small parameters $\varepsilon _{k}$ and only contains
terms that become time-independent in appropriate reference frames.

The resulting \textit{resonance expansion }in the leading order in each
effective coupling constant $g_{1}\epsilon _{\pm k}$ has the form

\begin{equation}
H_{RE\pm }^{F}\approx \tilde{\omega}X_{0}+\nu E_{0}+g_{1}\sum_{k=0}^{\infty
}\epsilon _{\pm k}\left( E^{k+1}X_{+}+E^{k+1}X_{-}\right) ,  \label{HeffG}
\end{equation}%
where $\tilde{\omega}$ is a modified $X$-system frequency, $\epsilon _{\pm
0}=1$,
\begin{equation}
\epsilon _{\pm k}=\frac{\mathrm{B}_{k}\left( a_{\pm 1},a_{\pm 2},\ldots
,a_{\pm k}\right) }{k!},  \label{e2n}
\end{equation}%
where
\begin{equation*}
a_{\pm k}=-\nu ^{-1}(k-1)!h_{\pm k},
\end{equation*}%
and $\mathrm{B}_{k}\left( a_{\pm 1},a_{\pm 2},\ldots ,a_{\pm k}\right) $ are
the complete Bell polynomials \cite{combinatoria}, and the constants $h_{\pm
k}$ are obtained recursively according to $h_{\pm 1}=g_{0}$, $h_{\pm 2}=\pm
2g_{1}^{2}/\left( \omega +\nu \right) $ and
\begin{eqnarray}
h_{\pm (2k+2)} &=&-\frac{h_{\pm (k+1)}^{2}}{2\left( \omega +(2k+1)\nu
\right) }  \label{h2k} \\
&&-\frac{1}{\omega +(2k+1)\nu }\sum_{m=1}^{k}h_{\pm m}h_{\pm (2k+2-m)},
\notag \\
h_{\pm (2k+1)} &=&-\frac{1}{\omega +2k\nu }\sum_{m=1}^{k}h_{\pm m}h_{\pm
(2k+1-m)},  \label{h2k+1}
\end{eqnarray}%
for $k=1,\ldots $. One can appreciate that $\epsilon _{\pm k}$ is a $k$-th
order homogeneous polynomial $\varepsilon ^{(k)}$ (\ref{epsilonk}) on some
small parameters $\varepsilon \sim g_{0,1}/l.c.(\omega ,\nu )\ll 1$, where $%
l.c.(\omega ,\nu )$ represents non zero linear combinations of $\omega $ and
$\nu $ for any relation between the frequencies.

The resonance expansion (\ref{HeffG}) contains all possible effective
resonant transitions that may emerge in (\ref{H1dop0}) and indicates that
such transitions happen only at $(k+1)\nu =\tilde{\omega}$, where the case $%
k=0$ corresponds to the principal resonance present in the original
Hamiltonian.

In principle, the effective Hamiltonian, describing the system excitation in
the vicinity of every particular resonance, should still be obtained from
the resonance expansion (\ref{HeffG}) by removing all of the other
resonances. However, as is proven in Appendix A, the elimination of all
terms in (\ref{HeffG}) that become non-resonant under the condition $\omega
\approx (k+1)\nu $ \textit{does not change} the leading order of the
effective interaction constants $g_{1}\epsilon _{\pm k}$, thus the effective
Hamiltonian has the form
\begin{eqnarray}
H_{\pm (k+1)}^{F} &\approx &\tilde{\omega}_{\pm (k+1)}X_{0}+\nu E_{0}  \notag
\\
&&+g_{1}\epsilon _{\pm k}\left( E^{k+1}X_{+}+E^{\dag k+1}X_{-}\right) .
\label{Hkeff}
\end{eqnarray}%
The effective $X$-system frequency $\tilde{\omega}_{\pm (k+1)}$ includes
small shifts that should be taken into account up to the order of the
coupling constant $g_{1}\epsilon _{\pm k}$, which determines the width of
the corresponding resonance.

The evolution operator corresponding to the effective Hamiltonian (\ref%
{Hkeff}) under the resonance condition $\tilde{\omega}_{\pm (k+1)}=(k+1)\nu $
is%
\begin{equation}
U_{\pm }^{F}(t)=e^{-i\nu t\left( (k+1)X_{0}+E_{0}\right)
}e^{-itg_{1}\epsilon _{\pm k}\left( E^{k+1}X_{+}+E^{\dag k+1}X_{-}\right) },
\label{UF}
\end{equation}%
and can be disentangled in the standard way. Using (\ref{UF}), the evolution
of any observable can be computed without returning to the time-dependent
frame. This is achieved by transforming the corresponding $X$-system
operator with (\ref{UF}) and averaging the result over the phase states (\ref%
{phaseSt}). Strictly speaking, the evolution operator should still be
transformed with all the transformations of the form (\ref{V}) used for
removing non-resonant terms in order to obtain the effective Hamiltonian (%
\ref{Hkeff}). Nevertheless, since the transformations (\ref{V}) are time
independent, they lead only to small modifications of amplitudes and can be
neglected in the first approximation. For instance, the evolution of $X_{0}$
operator in the resonance $\tilde{\omega}_{\pm (k+1)}=(k+1)\nu $ can be
easily found using (\ref{UF}),%
\begin{equation*}
X_{0}(t)=X_{0}C(2tg_{1}\epsilon _{\pm k})+\frac{i}{2}\left(
X_{-}-X_{+}\right) S(2tg_{1}\epsilon _{\pm k}),
\end{equation*}%
where $C(x)=\cos (x)$, $S(x)=\sin (x)$ for the $su(2)$ algebra and $%
C(x)=\cosh (x)$, $S(x)=\sinh (x)$ for the $su(1,1)$ algebra.

The frequency shifts for the lowest resonances can be easily found by a
direct application of the transformations given in Appendix A. In order to
obtain $\tilde{\omega}_{\pm (k+1)}$ for the highest order resonances the
following procedure can be applied: the effective Hamiltonian (\ref{Hkeff})
is unitary equivalent to (\ref{H1dop0}) up to $k$-th order on some small
parameters. In other words there exists a unitary transformation of the form
\begin{equation}
O_{\pm }=\exp \left\{ (\alpha _{\pm }^{\dag }-\alpha _{\pm })X_{0}+\beta
_{\pm }X_{+}-\beta _{\pm }^{\dag }X_{-}\right\} ,  \label{Uomega}
\end{equation}%
where
\begin{equation}
\alpha _{\pm }(E)=\sum_{j=1}a_{\pm j}E^{j},\quad \beta _{\pm }(E,E^{\dag
})=\sum_{j=0}b_{\pm j}E^{j}+c_{\pm j}E^{\dag j},  \label{abc}
\end{equation}%
such that under the condition $\omega \approx (k+1)\nu $
\begin{equation*}
O_{\pm }^{\dagger }H^{F}O_{\pm }=H_{\pm (k+1)}+\mathcal{\ O}(\varepsilon
^{(k+1)}).
\end{equation*}%
Taking into account the form of the perturbative action of transformations
of the type (\ref{Uomega}) on the Hamiltonian (\ref{H1dop0}), as discussed
in Appendix A, we realize that every coefficient in (\ref{abc}) can be
expanded in a series on some small parameters to be determined
\begin{equation}
x_{\pm j}=\sum_{m}x_{\pm j}^{(m)},\quad x_{\pm j}^{(m)}\sim \varepsilon
^{(j+2m)},  \label{xexp}
\end{equation}%
here $x_{\pm j}=a_{\pm j},b_{\pm j},c_{\pm j}$\textbf{$,$} except for $%
b_{\pm 0}=c_{\pm 0}=\sum b_{\pm 0}^{(m)}$, with $b_{\pm 0}^{(m)}\sim
\varepsilon ^{(2m+2)}$. \ In general,
\begin{eqnarray*}
O_{\pm }^{\dagger }H^{F}O_{\pm } &=&A_{\pm }\left( \alpha _{\pm },\alpha
_{\pm }^{\dag },\beta _{\pm },\beta _{\pm }^{\dag }\right) X_{0}+\nu E_{0} \\
&&+B_{\pm }\left( \alpha _{\pm },\alpha _{\pm }^{\dag },\beta _{\pm },\beta
_{\pm }^{\dag }\right) X_{+}+h.c.,
\end{eqnarray*}%
where the operators $A_{\pm }$ and $B_{\pm }$ can be easily found. Expanding
$A_{\pm }\left( \alpha _{\pm },\alpha _{\pm }^{\dag },\beta _{\pm },\beta
_{\pm }^{\dag }\right) $ and $B_{\pm }\left( \alpha _{\pm },\alpha _{\pm
}^{\dag },\beta _{\pm },\beta _{\pm }^{\dag }\right) $ in series according
to (\ref{xexp}) and equaling to $H_{\pm (k+1)}$ up to $\varepsilon ^{(k)}$
one can, in principle, determine all needed $x_{\pm j}^{(m)},m\leq k$ and
eventually find $\tilde{\omega}_{\pm (k+1)}$. However, such a procedure,
although systematic, faces significant numerical difficulties and in
practice is not very efficient.

\subsection{Examples}

\subsubsection{Semi-classical Rabi model}

The semi-classical Rabi model describes the evolution of an $S$-spin system
in a periodic field and the corresponding time-dependent Hamiltonian has the
form%
\begin{equation}
H_+(t)=\omega S_{z}+2g\left( S_{-}+S_{+}\right) \cos \nu t,  \label{HDicke}
\end{equation}%
where $S_{z,\pm }$ are generators of the $2S+1$ dimensional representation
of the $su(2)$ algebra, with
\begin{equation*}
\left[ S_{z},S_{\pm }\right] =\pm S_{\pm },\hspace{0.15in}\left[ S_{+},S_{-}%
\right] =2S_{z}.
\end{equation*}%
The Floquet form of (\ref{HDicke}) is
\begin{equation*}
H_+^{F}=\omega S_{z}+\nu E_{0}+g\left( E^{\dagger }S_{-}+h.c.\right)
+g\left( E^{\dagger }S_{+}+h.c.\right) ,
\end{equation*}%
where $E_{0},E^{\dagger },E$ are the generators of the Euclidian algebra (%
\ref{E}). The Hamiltonian (\ref{HDicke}) corresponds to $g_{0}=0$ in (\ref%
{Hint1d0}), so that $h_{+(2k+1)}=\epsilon _{+(2k+1)}=0$, and thus only odd
resonances in (\ref{HeffG}) survive,%
\begin{equation}
H_{RE+}^{F}=\tilde{\omega}S_{z}+\nu E_{0}+g\sum_{k=0}\epsilon _{+2k}\left(
E^{2k+1}S_{+}+h.c.\right) ,  \notag
\end{equation}%
where $\epsilon _{+2k}$ are given in (\ref{e2n}) and $\tilde{\omega}$ is the
shifted atomic frequency. In the vicinity of the resonance $\omega \approx
(2k+1)\nu $, the effective Hamiltonian takes the form
\begin{equation}
H_{+(2k+1)}^{F}\approx \tilde{\omega}_{+(2k+1)}S_{z}+\nu
E_{0}+g_{+(2k+1)}\left( E^{2k+1}S_{+}+h.c.\right) .  \notag
\end{equation}

The spin frequency shifts $\delta \omega _{+(2k+1)}=\tilde{\omega}%
_{+(2k+1)}-\omega +\mathcal{O}(\varepsilon ^{(2k+1)})$ and effective
couplings $g_{+(2k+1)}$ are given in Table \ref{firstSemiRabi} for $k=0,1,2$%

\begin{table}
\begin{center}
\begin{tabular}{|c|c|c|}
\hline
Interactions	& $g_{eff}$	& $\delta \omega _{+(2k+1)}$\\ \hline
$ES_++h.c. $ & $g$ & $g\varepsilon$ \\
$E^{3}S_++h.c. $ & $-\frac{9}{4}g\varepsilon^2$ & $\frac{9}{2}g\varepsilon +%
\mathcal{O}(\varepsilon^3)$ \\
$E^{5}S_++h.c. $ & $\frac{1}{9}\left(\frac{5}{2}\right)^5g\varepsilon^4$ & $%
\frac{25}{6}g\varepsilon\left(1-\frac{19}{144}\varepsilon^2\right)+\mathcal{O%
}(\varepsilon^5)$\\
\hline
\end{tabular}
\end{center}
\caption{The frequency shifts $\protect\delta \protect\omega _{+(2k+1)}=%
\tilde{\protect\omega}_{+(2k+1)}-\protect\omega +\mathcal{O}(\protect%
\varepsilon ^{(2k+1)})$ and effective couplings $g_{+(2k+1)},$ $k=0,1,2$ for
the semiclassical Rabi model in terms of the small parameter $\protect%
\varepsilon =g/\protect\omega $.}\label{firstSemiRabi}
\end{table}

\subsubsection{Quantum parametric oscillator}

For the quantum parametric oscillator,
\begin{equation*}
H_-(t)=\frac{p^{2}}{2}+\omega ^{2}\left( 1+2\gamma \cos \nu t\right) \frac{%
x^{2}}{2},
\end{equation*}%
corresponding to $g_{0}=g_{1}=g=\omega \gamma /2$, and $X_{+}\rightarrow
K_{+}=a^{\dag 2}/2$, $X_{-}\rightarrow K_{-}=a^{2}/2$, $X_{0}\rightarrow
K_{0}=\left( a^{\dag }a+aa^{\dag }\right) /2$ where , $K_{\pm },K_{0}$ are
generators of the $su(1,1)$ algebra, the expansion (\ref{Hint1d0}) is
reduced to
\begin{equation}
H_{RE-}^{F}=\tilde{\omega}a^{\dag }a+\nu E_{0}+\frac{g}{2}\sum_{k=0}^{\infty
}\epsilon _{-k}\left( a^{\dag 2}E^{k+1}+h.c.\right) ,  \notag
\end{equation}%
where $\epsilon _{-k}$ are given in (\ref{e2n}). As in the classical case,
the effective couplings corresponding to resonances at $2\omega \approx
(k+1)\nu $ are proportional to $\sim \omega \left( g/\omega \right) ^{k+1}$
and
\begin{equation*}
H_{-(k+1)}^{F}\approx \tilde{\omega}_{-(k+1)}a^{\dag }a+\nu
E_{0}+g_{-(k+1)}\left( E^{k+1}a^{\dag 2}+h.c.\right)
\end{equation*}

\begin{table}
\begin{center}
\begin{tabular}{|c|c|c|}
\hline
Interaction & $g_{eff}$ & $\delta \omega _{-(k+1)}$ \\ 
\hline
$Ea^{\dag 2}+h.c. $ & $\frac{1}{2}g$ & $-\frac{1}{4}g\varepsilon$ \\ 
$E^{2}a^{\dag 2}+h.c. $ & $-g\varepsilon$ & $-\frac{4}{3}g\varepsilon +%
\mathcal{O}(\varepsilon^3)$ \\ 
$E^{3}a^{\dag 2}+h.c. $ & $\frac{81}{32}g\varepsilon^2$ & $-\frac{9}{8}%
g\varepsilon+\mathcal{O}(\varepsilon^3)$ \\ 
\hline
\end{tabular}
\end{center}
\caption{The frequency shifts $\protect\delta \protect\omega _{-(k+1)}=%
\tilde{\protect\omega}_{-(k+1)}-\protect\omega +\mathcal{O}(\protect%
\varepsilon ^{(k+1)}) $ and effective couplings $g_{-(k+1)},$ $k=0,1,2$ for
the parametric quantum oscillator in terms of the small parameter $\protect%
\varepsilon =g/\protect\omega $.}\label{firstOscPar}
\end{table}

The oscillator frequency shifts $\delta \omega _{-(k+1)}=\tilde{\omega}%
_{-(k+1)}-\omega +\mathcal{O}(\varepsilon ^{(k+1)})$ and effective couplings
are given in Table \ref{firstOscPar} for $k=0,1,2$.

\subsection{Effective Hamiltonian, $g_{1}\ll g_{0}\sim \protect\omega ,%
\protect\nu $}

The situation is less involved in the limit $g_{1}\ll g_{0}\sim \omega ,\nu $
if the expansion of the effective coupling constants is restricted by the
leading order in the expansion of the effective coupling constants. Applying
the transformation%
\begin{equation}
V=\exp \left[ \frac{\epsilon }{2}\left( E^{\dag }-E\right) X_{0}\right] ,
\label{VM}
\end{equation}%
where $\epsilon =2g_{0}/\nu ,$ to the Hamiltonian (\ref{H1dop0}) the
following expression is obtained
\begin{eqnarray}
H_{\pm 0} &=&\omega X_{0}+\nu E_{0}-\frac{2g_{1}}{\epsilon }%
\sum_{k=1}(-1)^{k}kJ_{k}(\epsilon )\left( E^{k}X_{+}+E^{\dag k}X_{-}\right)
\notag \\
&&+\frac{2g_{1}}{\epsilon }\sum_{k=1}kJ_{k}(\epsilon )\left( E^{\dag
k}X_{+}+E^{k}X_{-}\right) ,  \label{H1sis0}
\end{eqnarray}%
where $J_{k}(\epsilon )$ are the Bessel functions. Removing all CR terms in (%
\ref{H1sis0}) in the weak interaction limit $g_{1}\ll \omega ,\nu $, results
in the following resonance expansion
\begin{eqnarray}
H_{\pm RE} &\approx &\omega X_{0}+I_{\pm }(\epsilon )X_{0}+\nu E_{0}  \notag
\\
&&-\frac{2g_{1}}{\epsilon }\sum_{k=1}(-1)^{k}kJ_{k}(\epsilon )\left(
E^{k}X_{+}+E^{\dag k}X_{-}\right)   \label{H1sisRE}
\end{eqnarray}%
where $\varepsilon _{k}=g_{1}/(\omega +k\nu )$ and
\begin{equation}
I_{\pm }(\epsilon )=\pm \frac{8g_{1}}{\epsilon ^{2}}\sum_{k=1}\varepsilon
_{k}k^{2}J_{k}^{2}(\epsilon ).
\end{equation}%
In the vecinity of each resonance $\omega \approx m\nu $, the effective
Hamiltonian takes the form
\begin{equation}
H_{eff\pm m}\approx \omega X_{0}+I_{\pm m}(\epsilon )X_{0}+\nu E_{0}+\frac{%
2g_{1}}{\epsilon }(-1)^{m+1}mJ_{m}(\epsilon )\left( E^{m}X_{+}+E^{\dag
m}X_{-}\right) ,
\end{equation}%
where the frequency corrections,
\begin{equation}
I_{\pm m}(\epsilon )=\pm \frac{16g_{1}^{2}\omega }{\epsilon ^{2}}\sum
_{\substack{ k=1 \\ k\neq m}}\frac{k^{2}J_{k}^{2}(\epsilon )}{\omega
^{2}-k^{2}\nu ^{2}}\pm \frac{8g_{1}^{2}m^{2}J_{m}^{2}(\epsilon )}{\epsilon
^{2}(\omega +m\nu )}.
\end{equation}%
appear as a result of eliminating all the other transitions in (\ref{H1sisRE}%
).

\subsection{Modulated quantum system with intensity dependent coupling}

Our approach can be easily extended to Hamiltonians non-linear on the
algebra generators when only the frequency of the system is modulated. Let
us consider the following Hamiltonian
\begin{eqnarray}
H_{\pm }(t) &=&\omega \left[ 1+\gamma \cos (\nu t)\right] X_{0}+H_{int}
\label{Hnl} \\
H_{int} &=&g\left[ X_{+}f(X_{0})+f(X_{0})X_{-}\right] ,  \label{Hint}
\end{eqnarray}%
where $f(X_{0})$ is a function of the \textquotedblleft
diagonal\textquotedblright\ operator $X_{0}$, in the strong modulation
limit, $\omega \gamma \lesssim \nu $. The interaction Hamiltonian in (\ref%
{Hint}) describes a wide class of quantum optical systems as atom-photon
interactions, parametric processes, etc \cite{Karas}. It is clear, that only
assisted transitions, i.e. induced by the external field, can be generated
by (\ref{Hnl}) due to the presence of the term $\omega X_{0}$.

Applying the transformation
\begin{equation}
V=\exp \left[ \frac{\epsilon }{2}\left( E^{\dag }-E\right) X_{0}\right] ,
\label{TransA}
\end{equation}%
where $\epsilon =\omega \gamma /\nu \lesssim 1$ to the Flouquet Hamiltonian
corresponding to (\ref{Hnl})
\begin{equation}
H_{\pm }^{F}=\omega X_{0}+\nu E_{0}+\frac{1}{2}\omega \gamma \left(
E+E^{\dag }\right) X_{0}+g\left[ X_{+}f(X_{0})+f(X_{0})X_{-}\right] ,
\end{equation}%
we obtain the following exact expression
\begin{eqnarray}
VH_{\pm }^{F}V^{\dagger } &=&\omega X_{0}+\nu E_{0}+gJ_{0}(\epsilon )\left[
X_{+}f(X_{0})+f(X_{0})X_{-}\right]   \notag \\
&&+g\sum_{k=1}J_{k}(\epsilon )\left[ \left( E^{\dag k}+(-1)^{k}E^{k}\right)
X_{+}f(X_{0})+h.c.\right] .  \label{HNL11}
\end{eqnarray}%
The Hamiltonian (\ref{HNL11}) contains all the possible resonances $\sim
E^{k}X_{+}f(X_{0})$, along with CR terms $\sim E^{\dag k}X_{+}f(X_{0})$,
which can be perturbatively eliminated in the weak couppling limit, $g\ll
\omega ,\nu $, by a set of transformations
\begin{equation}
W_{m}=\exp \left( \varepsilon _{m}J_{m}(\epsilon )\left[ E^{\dag
m}X_{+}f(X_{0})-f(X_{0})E^{m}X_{-}\right] \right) ,  \notag
\end{equation}%
where $\varepsilon _{m}=g/(\omega +m\nu )\ll 1$. This leads to corrections
of order $\varepsilon ^{(1)}\sim g/l.c.(\omega ,\nu )$ in the frequency, and
$\varepsilon ^{(2)}\sim \left( \varepsilon ^{(1)}\right) ^{2}$ in the
coupling constant, and, in addition, to new CR terms of the form $%
\varepsilon ^{(1)}(E^{\dag l}+h.c.)K_{\pm }(X_{0}))$, where%
\begin{equation*}
K_{\pm }(X_{0})=\pm 2X_{0}f^{2}(X_{0})+X_{+}X_{-}\left(
f^{2}(X_{0}-1)-f^{2}(X_{0})\right) ,
\end{equation*}%
which can be also removed with an appropriate transformation. As a result we
arrive at the following resonant expansion%
\begin{eqnarray}
H_{RE\pm }^{F} &\approx &\tilde{\omega}X_{0}+\nu E_{0}+gI(\epsilon )K_{\pm
}(X_{0})  \notag \\
&&+g\sum_{k=1}^{\infty }(-1)^{k}J_{k}(\epsilon )\left(
E^{k}X_{+}f(X_{0})+h.c.\right) ,  \label{HNL12}
\end{eqnarray}%
where $\tilde{\omega}=\omega (1+O(\varepsilon ^{(1)}))$ and
\begin{equation}
I(\epsilon )=g\sum_{k=0}^{\infty }\frac{J_{k}^{2}(\epsilon )}{\omega +k\nu }.
\label{I}
\end{equation}%
The expansion (\ref{HNL12}) is similar to (\ref{HeffG}), exhibiting possible
effective resonances at $\omega \approx k\nu $. However, the non-linear term
$\sim g\varepsilon ^{(1)}K_{\pm }(X_{0})$ generates an intensity dependent
frequency shift, which makes the resonances with $k\gtrsim M$, where $%
J_{M}(\epsilon )\sim I(\epsilon )$ inefficient.

It is easy to find that the effective Hamiltonian in the vecinity of $m$-th
resonace, $\omega \approx m\nu $, $m<M$, has the form
\begin{eqnarray}
H_{\pm m}^{F} &\approx &\tilde{\omega}X_{0}+\nu E_{0}+g\tilde{I}%
_{m}(\epsilon )K_{\pm }(X_{0})  \notag \\
&&+(-1)^{m}gJ_{m}(\epsilon )\left( E^{m}X_{+}f(X_{0})+h.c.\right) ,
\label{NLHE}
\end{eqnarray}%
where
\begin{equation}
\tilde{I}_{m}(\epsilon )=\frac{gJ_0^2(\epsilon)}{\omega}+2g\omega \sum
_{\substack{ k=1  \\ k\neq m}}^{\infty }\frac{J_{k}^{2}(\epsilon )}{\omega
^{2}-k^{2}\nu ^{2}}+g\frac{J_{m}(\epsilon )}{\omega +m\nu }.
\end{equation}

It is worth noting that in the weak modulation limit, $\epsilon \lesssim
\varepsilon \ll 1$, only the first resonance $\omega \approx \nu $ survives
in the non-linear case, since the effective coupling is of order of the
intensity dependent shift,
\begin{equation}
H_{\pm 1}^{F}\approx \omega X_{0}+\nu E_{0}+\frac{g^{2}}{\omega }K_{\pm
}(X_{0})-\frac{g\omega\gamma}{2\nu }\left( EX_{+}f(X_{0})+f(X_{0})E^{\dag
}X_{-}\right) .
\end{equation}

Observe, that in the particular case, $f(X_{0})=1$, the resonant expansion
for linear Hamiltonians is recovered. For instance, for the Dicke model
\begin{equation}
H=\omega _{0}\left[ 1+\gamma \cos (\nu t)\right] S_{z}+g\left(
S_{+}+S_{-}\right) .  \label{HD}
\end{equation}%
in the strong modulation limit, $\omega _{0}\gamma \gg g$, the resonace
expansion has the following form
\begin{equation}
H_{RE}^{F}\approx \omega _{0}S_{z}+2\frac{g^{2}}{\omega }J_{0}^{2}(\epsilon
)S_{z}+g\sum_{k=1}(-1)^{k}J_{k}(\epsilon )\left( E^{k}S_{+}+h.c.\right) .
\notag
\end{equation}

\section{Two periodically modulated coupled quantum systems}

The application of Lie transformations in order to determine the effective
interaction constants, corresponding to effective resonances emerging in the
case of two coupled and periodically modulated systems, becomes a quite
involved task. We will analyze the situation where the coupling between the
systems is significantly smaller than the bare frequencies of both systems.
Thus, for consistency, all CR terms in the interaction Hamiltonian,
appearing even in the absence of time-dependence, should be taken into
account.

Let us consider two interacting quantum systems $X$ and $Y$ in dipole
approximation, where the frequency of one of those is harmonically
modulated. The corresponding Hamiltonian is
\begin{equation}
H(t)=\omega _{0}\left[ 1+\gamma \cos (\nu t)\right] X_{0}+\omega
_{1}Y_{0}+g\left( X_{+}+X_{-}\right) \left( Y_{+}+Y_{-}\right) ,
\label{H2sistgent}
\end{equation}%
where the operators describing $X$ or $Y$ systems can be from $su(2)$, $%
su(1,1)$ or $h(1)$ algebras. The commutation relations have the following
generic form
\begin{equation}
\lbrack X_{+},X_{-}]=P_{x}(X_{0})=\nabla _{x}\phi _{x}(X_{0}),\qquad \lbrack
Y_{+},Y_{-}]=P_{y}(Y_{0})=\nabla _{y}\phi _{y}(Y_{0}),
\end{equation}%
where $\phi _{z}(Z_{0})=Z_{+}Z_{-}$ is a second degree polynomial for $su(2)$
and $su(1,1)$ algebras, and is a first degree polynomial for the
Heisenberg-Weyl algebra $h(1)$; the discrete derivative is defined as
\begin{equation}
\nabla _{nz}\phi _{z}(Z_{0})=\phi _{z}(Z_{0})-\phi _{z}(Z_{0}+n)  \notag
\end{equation}%
where $n\in \mathbb{Z}$.

The resonance expansion in the limit of strong modulation, $\omega \gamma
\lesssim \nu $ and weak coupling, $g\ll \omega _{0,1}$ is obtained in
Appendix \ref{B} and has a generic form%
\begin{equation}
H_{RE}^{F}\approx \omega _{0}X_{0}+\omega _{1}Y_{0}+\nu
E_{0}+gK(X_{0},Y_{0})+H_{int},  \label{RExy}
\end{equation}

The intensity dependent frequency shift $K(X_{0},Y_{0})$ explicitly given in
(\ref{P2gen})-(\ref{P33gen}), leads to inhibition of higher-order
transitions in $X$ and $Y$ systems. For the considered symmetries $h(1)$, $%
su(2)$ and $su(1,1)$, the effective interacion Hamiltonian $H_{int}$ has the
following structure%
\begin{equation*}
H_{int}=\sum_{k=1}\sum_{m=0}^{3}\sum_{n=1}\varepsilon
_{nk}^{(m)}H_{nk}^{(m)},
\end{equation*}%
where $H_{nk}^{(m)}$ are given in Appendix B.

The form of the intensity dependent shift $K(X_{0},Y_{0})$ (\ref{P2gen})-(%
\ref{P33gen}) depends on the degree of the polynomials $\phi _{x}(X_{0})$
and $\phi _{y}(Y_{0})$:

i) both $X$ and $Y$ systems are described by the $h(1)$ algebra. In this
case $K(X_{0},Y_{0})$ is a linear form on $X_{0},Y_{0}$.

ii) one of the systems is described by $h(1)$ and another by $su(2)/su(1,1)$
algebra. In this case the leading term in $K(X_{0},Y_{0})$ is a second
degree polynomial on $X_{0}$ and $Y_{0}$\textbf{,} and the first correction
is of a third degree one.

iii) The leading term in $K(X_{0},Y_{0})$ is a third degree polynomial if
both systems have $su(2)/su(1,1)$ symmetry.

\subsection{Modulated quantum parametric amplifier}

Let us start with a non-degenerated parametric quantum amplifier with
modulated interaction constant \cite{devoret}, described by
\begin{equation}
H(t)=\omega _{a}\left( 1+\gamma \cos \left( \nu t\right) \right) a^{\dag
}a+\omega _{b}b^{\dag }b+g\left( a^{\dag }+a\right) \left( b^{\dag
}+b\right) .  \label{Hap}
\end{equation}%
In this case $X_{+}=a^{\dag },~X_{-}=a$, $X_{0}=a^{\dag }a$, $Y_{+}=b^{\dag
},~Y_{-}=b$, $Y_{0}=b^{\dagger }b$ and no intensity dependent shift (\ref%
{P2gen})-(\ref{P33gen}) appears in the resonant expansion
\begin{equation*}
K(a^{\dag }a,b^{\dag }b)\approx -g\sum_{k=0}\varepsilon
_{k}J_{k}^{2}(\epsilon )\left( a^{\dag }a+b^{\dag }b+1\right) +O(\varepsilon
^{(3)}),\quad \varepsilon _{k}=\frac{g}{\omega _{a}+\omega _{b}+k\nu }
\end{equation*}%
since $\phi (a^{\dag }a)=a^{\dag }a$ and $\nabla \phi (a^{\dag }a)=1$. The
resonance expansion (\ref{RExy}) is reduced to the following
\begin{eqnarray}
H_{RE}^{F} &\approx &\left( \omega _{a}-gI_{a}(\epsilon )\right) a^{\dag
}a+\left( \omega _{b}-gI_{b}(\epsilon )\right) b^{\dag }b  \label{reampli} \\
&&+gJ_{0}(\epsilon )\left( ab^{\dag }+h.c.\right) +g\sum_{k=1}J_{k}(\epsilon
)\left[ \left( E^{\dag k}+(-1)^{k}E^{k}\right) ab^{\dag }+h.c.\right]  \notag
\\
&&+g\sum_{k=1}J_{k}(\epsilon )\left( E^{\dag k}ab+h.c.\right)
+g\sum_{k=1}\left( \varepsilon _{1k}^{(1)}(\epsilon )E^{\dag
k}a^{2}+\varepsilon _{2k}^{(1)}(\epsilon )E^{\dag k}b^{2}+h.c.\right) ,
\notag
\end{eqnarray}%
where $I(\epsilon )$ is defined in (\ref{I}) with $\omega =\omega
_{a}+\omega _{b}$ and
\begin{equation*}
\varepsilon _{1k}^{(1)}(\epsilon )\approx -g\sum_{l=0}\frac{%
(-1)^{k+l}J_{l}(\epsilon )J_{l+k}(\epsilon )}{\omega _{a}+\omega _{b}+l\nu }%
,\quad \varepsilon _{2k}^{(1)}(\epsilon )\approx -g\sum_{l=0}\frac{%
J_{l}(\epsilon )J_{l+k}(\epsilon )}{\omega _{a}+\omega _{b}+l\nu }.
\end{equation*}%
The effective two-photon resonances have significantly smaller width than
the assisted transitions already present in the Hamiltonian (\ref{Hap}), $%
ab^{\dag },$ $ab$. The effective Hamiltonians describing the principal, $%
\sim ab^{\dag }$ and side-band $\sim E^{\dag k}ab^{\dag },$ $E^{k}ab^{\dag
}, $ $E^{\dag k}ab$ transitions acquire frequency correction of order $%
O(\varepsilon ^{(2)})$ in the vicinity of each resonance. However, the
frequency shift and effective coupling constant in the vicinity of
two-photon transitions $2\omega _{a,b}\approx k\nu $ (obtained by removing
all the other resonances) are significantly modified. The frequency shift
takes the form
\begin{equation}
\tilde{I}_{a,b}(\epsilon )=\frac{2g\omega _{b,a}J_{0}^{2}(\epsilon )}{\omega
_{a}^{2}-\omega _{b}^{2}}+2g\sum_{n=1}\frac{J_{n}^{2}(\epsilon )\left(
\omega _{a}+\omega _{b}\right) }{\left( \omega _{a}+\omega _{b}\right)
^{2}-n^{2}\nu ^{2}}+2g\sum_{n=1}\frac{J_{n}^{2}(\epsilon )\left( \omega
_{b,a}-\omega _{a,b}\right) }{\left( \omega _{a}-\omega _{b}\right)
^{2}-n^{2}\nu ^{2}},  \label{Iampli}
\end{equation}%
where the values of the summation index satisfying $|\omega _{a}\pm \omega
_{b}|=n\nu $ are excluded.

For instance, in the case $2\omega _{b}\approx \nu $ y $\omega _{a}=\omega
_{b}/2$, the effective Hamiltonian
\begin{equation}
H_{eff}\approx \omega _{a}(1-\tilde{I}_{a}(\epsilon ))a^{\dag }a+\omega
_{b}(1-\tilde{I}_{b}(\epsilon ))b^{\dag }b+\nu E_{0}+g_{eff}\left( Eb^{\dag
2}+E^{\dag }b^{2}\right) ,  \label{Habt}
\end{equation}%
where\textbf{\ }
\begin{equation}
g_{eff}\approx 2g^{2}\omega _{b}\sum_{l=0}^{\infty }(2l+1)\frac{%
J_{l}(\epsilon )J_{l+1}(\epsilon )}{\omega _{a}^{2}-(\omega _{b}+l\nu )^{2}},
\label{h2fotones}
\end{equation}%
describes the $b$-mode squeezing.

\begin{figure}
\includegraphics[width=0.5\textwidth]{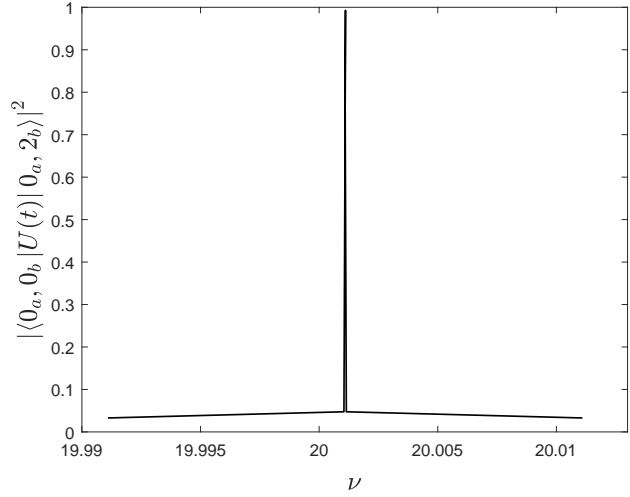}
\caption{Time averaged transition probability $|\langle
0_{a},0_{b}|U(t)|0_{a},2_{b}\rangle |^{2}$ as a function of the modulation
frequency $\protect\nu $ generated by the time-dependent Hamiltonian (%
\protect\ref{Hap}); $\protect\nu \approx 2\protect\omega _{b}$, $\protect%
\omega _{b}=10$, $\protect\omega _{a}=\protect\omega _{b}/2$, $g=0.1$ and $%
\protect\epsilon =0.9$.}
\label{Rampli}
\end{figure}

In Fig. \ref{Rampli}, we plot the time-averaged transition probability $%
|\langle 0_{a},0_{b}|U(t)|0_{a},2_{b}\rangle |^{2}$, where, for $g=0.1$, $%
\omega _{b}=10$ and $\epsilon =0.9$, the value $\nu =\omega _{b}-\tilde{I}%
_{b}\approx 20.0011$ is obtained, which perfectly coincides with the
numerical calculations.

\begin{figure}
\includegraphics[width=0.5\textwidth]{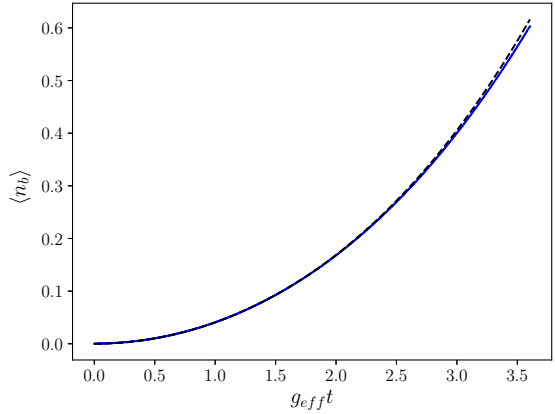}
\caption{Evolution of the avarage photon number in the mode $b$ for the
initial vacuum state in both modes in case of effective two-photon
transition, $\protect\nu \approx 2\protect\omega _{b}$, where $\protect%
\omega _{b}=10$, $\protect\omega _{a}=\protect\omega _{b}/2$, $g=0.1$ and $%
\protect\epsilon =0.9$. The solid (blue) line corresponds to the analytic
approximation (\protect\ref{nb}), the dashed (black) line results from
numerical calculation with the Hamiltonian (\protect\ref{Hap}).}
\label{fig2fotones}
\end{figure}

In Fig. \ref{fig2fotones} we compare the exact evolution of the average
photon number in the $b$-mode, starting with the initial vacuum state $%
|0_{a},0_{b}\rangle $ with the results of analytical calculations using the
effective Hamiltonian (\ref{Habt}),%
\begin{equation}
\langle b^{\dag }b\rangle_{app} \approx \sinh ^{2}\left( 2g_{eff}t\right) .
\label{nb}
\end{equation}

\subsubsection{Dicke model with modulated frequency}

The dynamics of the quantum Dicke model, describing the interaction between
an effective $S$-spin system and a single mode of a quantized field with
harmonically modulated atomic frequency \cite%
{cr,SQmicrowave,1atomLaser,casimir,sidebands} is governed by the following
Hamiltonian,
\begin{equation}
H=\omega _{0}\left[ 1+\gamma \cos \left( \nu t\right) \right] S_{z}+\omega
_{1}a^{\dag }a+g\left( a^{\dag }+a\right) \left( S_{+}+S_{-}\right) ,
\label{HDt}
\end{equation}%
where $g\ll \omega _{0,1}$ and, $0<\epsilon =\omega _{0}\gamma /\nu <1$,
which corresponds to (\ref{H2sistgent}) with $X_{\pm }=S_{\pm }$, $%
X_{0}=S_{z}$ and $Y_{+}=a^{\dag },~Y_{-}=a$, $Y_{0}=a^{\dag }a$. In this
case the intensity-dependent shift (\ref{P2gen})-(\ref{P33gen}) takes the
form:
\begin{equation*}
K(S_{z},a^{\dag }a)=g\varepsilon ^{(1)}(\epsilon )\left( S_{z}^{2}+\left(
1+2a^{\dag }a\right) S_{z}\right) +\varepsilon ^{(3)}(\epsilon )(a^{\dag
}a)^{2}S_{z}+O\left( \varepsilon ^{(3)}\right) ,
\end{equation*}%
where $\varepsilon ^{(k)}\sim \varepsilon ^{k}$, $\varepsilon \sim
g/l.c.(\omega ,\nu )\ll 1$, are some homogeneous polynomials on the Bessel
functions $J_{k}(\epsilon )$, $0<\epsilon =\omega _{0}\gamma /\nu \lesssim 1$
(\ref{epsilonkJ}). In particular, i) the dynamic Stark shift term $\sim
\varepsilon _{k}^{(1)}(\epsilon )a^{\dag }aS_{z}$ suppressess all
transitions between the field and the atomic system with an exchange of more
than one excitation; ii) the atomic Kerr term $\sim \varepsilon
_{k}^{(1)}(\epsilon )S_{z}^{2}$ does not allow to efficiently absorb more
than one excitation by the atomic system; iii) the field Kerr term $\sim
\varepsilon _{k}^{(3)}(\epsilon )(a^{\dag }a)^{2}$ makes the generation of
more than four photons by the quantum field inefficient. Thus, the resonance
expansion containing only efficient transitions takes the form
\begin{eqnarray}
H_{RE}^{F} &\approx &\omega _{0}S_{z}+\omega _{1}a^{\dag }a+\nu
E_{0}+K(S_{z},a^{\dag }a)+gJ_{0}(\epsilon )\left( aS_{+}+h.c.\right)   \notag
\\
&&+g\sum_{k=1}J_{k}(\epsilon )\left[ \left( E^{\dag k}+(-1)^{k}E^{k}\right)
aS_{+}+E^{k}a^{\dag }S_{+}+h.c.\right]   \notag \\
&&+g\sum_{k=1}\varepsilon _{1k}^{(1)}(\epsilon )\left(
E^{k}S_{+}^{2}+h.c.\right)   \notag \\
&&+g\sum_{k=1}\varepsilon _{2k}^{(1)}(\epsilon )\left( E^{k}a^{\dag
2}S_{z}+h.c.\right) +g\sum_{k=1}\varepsilon _{3k}^{(3)}(\epsilon )\left(
a^{\dag 4}S_{z}+h.c.\right) .  \label{Dicke}
\end{eqnarray}

\subsubsection{Non-symmetric excitation of an atomic system in a vacuum field}

The resonant expansion (\ref{Dicke}) reveals the existence of effective
processes consisting in the excitation of two atoms in the symmetric
configuration, described by
\begin{equation}
g\varepsilon_{1k} ^{(1)}(\epsilon) \left( E^kS_{+}^{2}+h.c.\right) .  \notag
\end{equation}%
However, this process is rapidly suppressed by the atomic Kerr term $\sim
S_{z}^{2}$, which is of the first order on the small parameters. Thus, the
symmetry of the atomic system should be broken in order to render the
two-atom excitation process efficient.

Let us consider the following generalization of the Hamiltonian (\ref{HDt})
to the two-atom case,
\begin{eqnarray}
H(t) &=&\sum_{i=1}^{2}\left( \omega _{i}+g_{0}\cos \left( \nu t\right)
\right) s_{zi}+\omega _{c}a^{\dag }a  \notag \\
&&+2\left( a^{\dag }+a\right) \sum_{i=1}^{2}g_{i}s_{xi},  \label{H2At}
\end{eqnarray}%
which corresponding Flouquet form is
\begin{eqnarray}
H^{F} &=&\sum_{i=1}^{2}\omega _{i}s_{zi}+\omega _{c}a^{\dag }a+\nu E_{0}
\label{DMda} \\
&&+g_{0}\left( E+E^{\dag }\right) \sum_{i=1}^{2}s_{zi}+2\left( a^{\dag
}+a\right) \sum_{i=1}^{2}g_{i}s_{xi}.  \notag
\end{eqnarray}%
For simplicity, we also assume that $\epsilon =g_{0}/\nu \ll 1$. In this
case we obtain a resonant expansion, which up to the second order on the
small parameters is given in Appendix \ref{b1}, Eq. (\ref{2atom}). Instead,
the Kerr term Eq. (\ref{2atom}) contains the spin exchange operator $%
s_{-1}s_{+2}+h.c.$, which can be taken out of resonance under appropriate
frequency conditions. For instance, choosing
\begin{eqnarray}
\omega _{c} &=&\omega _{1}+\omega _{2}\approx \nu ,  \notag \\
g_{i} &\ll &|\omega _{1}-\omega _{2}|,  \notag
\end{eqnarray}%
Imposing appropriate conditions on the frequencies all of the first order
transitions in (\ref{2atom}) can be removed thus arriving at the following
effective Hamiltonian for the initial vacuum field mode
\begin{equation}
H_{eff}^{F}\approx \tilde{\omega}_{1}s_{z1}+\tilde{\omega}_{2}s_{z2}+\nu
E_{0}+g_{eff}\left( Es_{+1}s_{+2}+h.c.\right) ,  \label{He2at}
\end{equation}%
where $\tilde{\omega _{i}}=\omega _{i}+g_{i}\left( \varepsilon _{1i}+\delta
_{1i}\right) +\mathcal{O}(\varepsilon ^{(3)})$, $\epsilon =g_{0}/\nu $, $%
\varepsilon _{1i}=g_{i}/(\omega _{c}+\omega _{i})$, $\delta
_{1i}=g_{i}/(\omega _{i}-\omega _{c})$, and $\varepsilon _{2i}=g_{i}/\omega
_{c}$. The effective interaction constant is
\begin{equation}
g_{eff}=\epsilon \left( g_{1}\varepsilon _{12}+g_{2}\varepsilon
_{11}-g_{1}\delta _{12}-g_{2}\delta _{11}\right) .  \notag
\end{equation}%
The Hamiltonian (\ref{He2at}) describes an effective excitation of two
different atoms mediated by a vacuum field in the modulated Dicke model \cite%
{dodonov2}.

\begin{figure}
\includegraphics[width=0.5\textwidth]{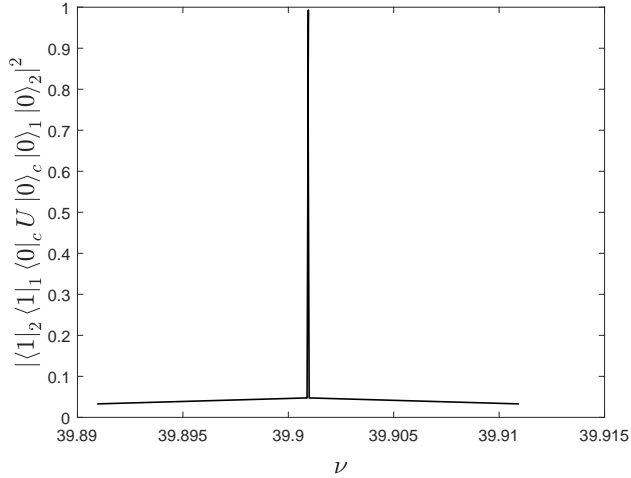}
\caption{Time averaged transition probability $|\langle
0_{1},0_{2}|U(t)|1_{1},1_{2}\rangle |^{2}$ as a function of the modulation
frequency $\protect\nu $ generated by the time-dependent Hamiltonian (%
\protect\ref{H2At}); $\protect\nu \approx \protect\omega _{1}+\protect\omega %
_{2}$, $\protect\omega _{1}=10$, $\protect\omega _{2}=3\protect\omega _{1}$,
and $\protect\omega _{c}=\protect\omega _{1}+\protect\omega _{2}$ with $%
g_{0}=g_{1}=g_{2}=1$.}
\label{trans2A}
\end{figure}

\begin{figure}
\includegraphics[width=0.45\textwidth]{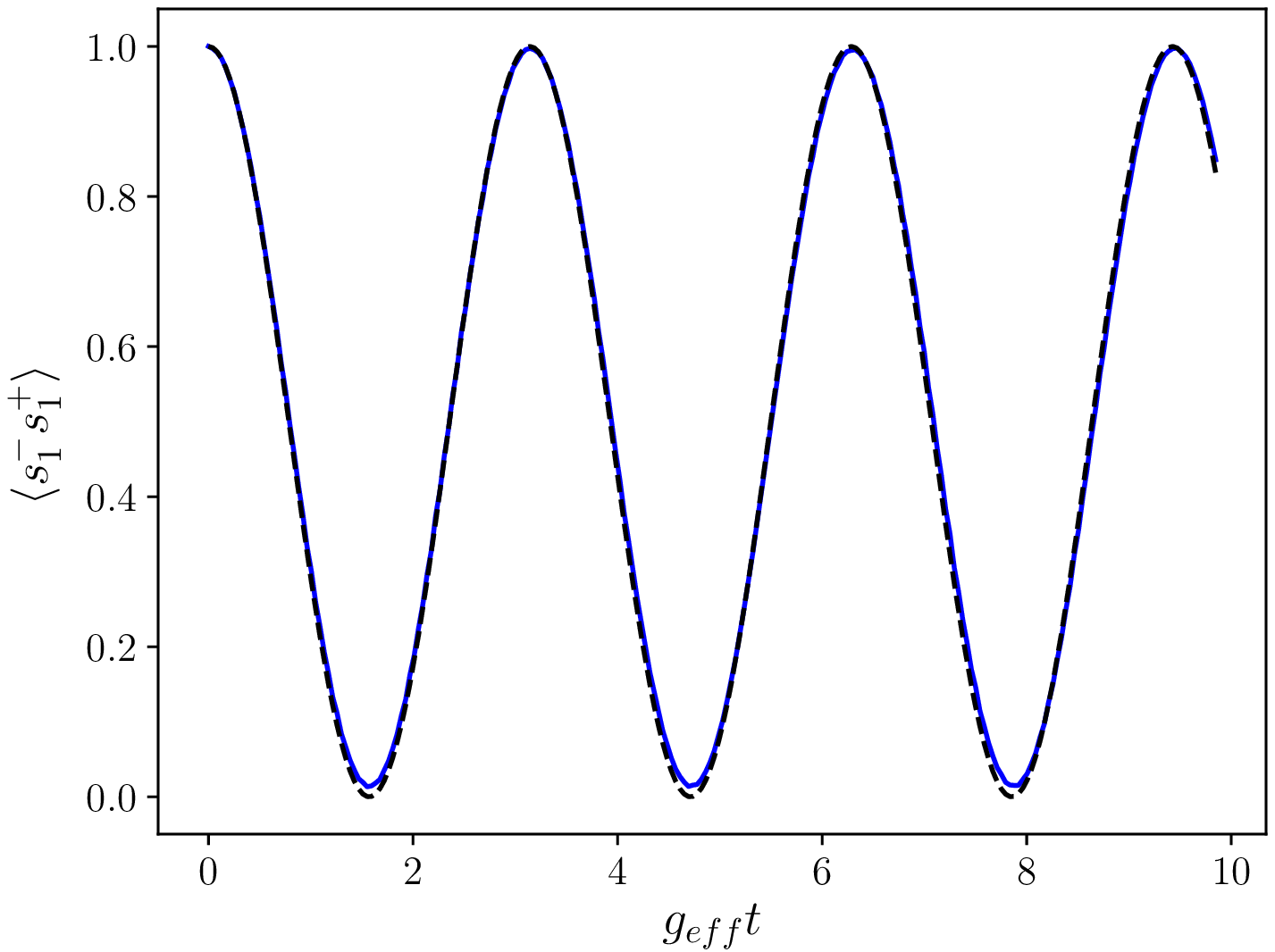} \includegraphics[width=0.45\textwidth]{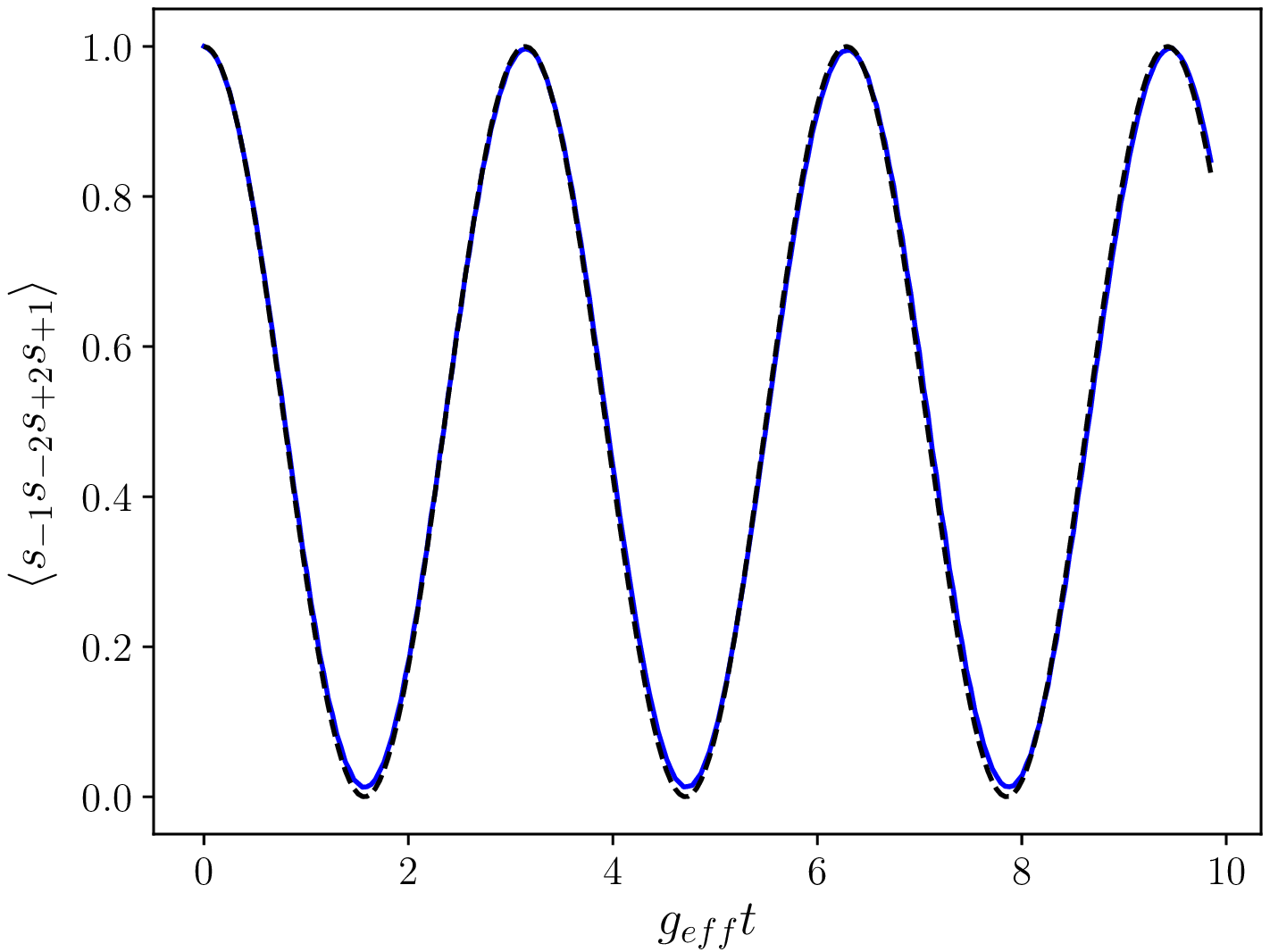}
\caption{Evolution of the averages $\langle s_{-1}s_{+1}\rangle $ (above)
and $\langle s_{-1}s_{-2}s_{+2}s_{+1}\rangle $ (below) for the initial
non-excited atoms and the cavity mode in vacuum $|0_{1},0_{2}\rangle $; $%
\protect\nu =\tilde{\protect\omega}_{1}+\tilde{\protect\omega}_{2}$, $%
\protect\omega _{2}=3\protect\omega _{1}$, $\protect\omega _{c}=\protect%
\omega _{1}+\protect\omega _{2}$ with $\protect\omega _{1}=10$, and $%
g_{0}=g_{1}=g_{2}=1$; the continuous (blue) line corresponds to the
numerical calculation for the exact Hamiltonian (\protect\ref{H2At}), the
dashed (black) line corresponds to the approximation (\protect\ref{Ev2at}).}
\label{2atomos}
\end{figure}

In Fig. \ref{trans2A}, we plot the time-averaged probability of two-atom
excitation $|\langle 0_{1},0_{2}|U(t)|1_{1},1_{2}\rangle |^{2}$ from the
vacuum as a function of the modulation frequency $\nu $, at $\omega
_{2}=3\omega _{1}$ with $\omega _{1}=10$, $g_{0}=g_{1}=g_{2}=g=1$. The
evolution operator $U(t)$ is generated by the exact Hamiltonian (\ref{H2At}%
). The position of the resonance is well described by our approximation,
giving $\nu =\tilde{\omega}_{1}+\tilde{\omega}_{2}\approx 4\omega
_{1}-104g/(105\omega _{1})=39.90095$.

In figure \ref{2atomos} we compare the evolution of averages $\langle
s_{-1}s_{+1}\rangle $ and $\langle s_{-1}s_{-2}s_{+2}s_{+1}\rangle $,
describing the excitation of the first atom and the joint excitation of both
atoms \cite{1f2a}, and the corresponding approximate evolutions, generated
by the effective Hamiltonian (\ref{He2at}), for the initial non-excited
atoms and the cavity mode in vacuum $|0_{1},0_{2}\rangle $. The approximate
expressions, immediately following from (\ref{He2at}),
\begin{equation}
\langle s_{-1}s_{+1}\rangle _{app}=\langle s_{-1}s_{-2}s_{+2}s_{+1}\rangle
_{app}\approx \cos ^{2}\left( g_{eff}t\right) ,  \label{Ev2at}
\end{equation}%
describe the dynamics of the observables fairly well for $\nu =\tilde{\omega}%
_{1}+\tilde{\omega}_{2}$, $\omega _{2}=3\omega _{1}$, $\omega _{c}=\omega
_{1}+\omega _{2}$ with $\omega _{1}=10$, and $g_{0}=g_{1}=g_{2}=g=1$.

\section{Conclusions}

Even the simplest periodically modulated quantum systems exhibit a rich
resonance structure captured by the expansion (\ref{HeffG}). This resonance
expansion is obtained by a specific Lie-type perturbation theory where
coupling constant is small with respect to the bare system%
\'{}%
s frequencies both for weak and strong modulation amplitude. In the
framework of this approach the order of each resonance, which determines the
width of the related transition, and consequently the Rabi frequency of
corresponding oscillations can be found. In case of single modulated linear
systems we have been able to obtain the principal contribution to the
effective interaction constant corresponding to each resonant term.

Effective Hamiltonians, describing all possible resonant transitions, can be
extracted from the resonance expansion by establishing some particular
frequency conditions. It was observed that in the case of a single modulated
system the order of effective Hamiltonians in the vicinity of each resonance
is exactly the same as that of the corresponding terms in the resonance
expansion.

The common feature of modulated linear (on Lie algebra generators, in our
case $su(2)$ and $su(1,1)$) Hamiltonians is the absence of the dynamic Stark
shift and Kerr-like terms in the resonance expansion. Thus, all of the
resonances appearing in this expansion are \textit{efficient} i.e., there
are always frequency conditions such that the transition probabilities
between energy levels, described by the corresponding effective Hamiltonian,
are close to unity. In contrast, the effective Hamiltonians of periodically
perturbed non-linear quantum systems (effective as in Sec. 3.3) and coupled
(as in Sec. 4), always contain non-linearities that \textquotedblleft
select\textquotedblright\ the efficient transitions among all those present
in the formal resonance expansion.

\appendix

\section{Single periodically modulated quantum system}

\label{A}

Here we obtain the resonance expansion corresponding to the Flouquet
Hamiltonian (\ref{H1dop0}) by removing CR terms with adequate small Lie
transformations and keeping only the principal order on $g_{0,1}\ll \omega
,\nu $.

The CR term $g_{1}\left( E^{\dag }X_{+}+h.c.\right) $ can be exactly
eliminated by the transformation
\begin{equation}
V_{\pm 1}=\exp \left\{ \varepsilon _{\pm 1}\left( E^{\dag
}X_{+}-EX_{-}\right) \right\} ,  \label{U11d0}
\end{equation}%
where
\begin{equation}
T\left( 2\varepsilon _{\pm 1}\right) =\frac{2g_{1}}{\omega +\nu },  \label{t}
\end{equation}%
and $T(x)=\tan (x)$ for $su(2)$ case, $[X_{+},X_{-}]=2X_{0}$, and $%
T(x)=\tanh (x)$ for the case $su(1,1)$, $[X_{+},X_{-}]=-2X_{0}$. The
Hamiltonian (\ref{H1dop0}) transformed with $V_{\pm 1}$ takes the form $%
H_{\pm 1}=V_{\pm 1}H_{\pm }^{F}V_{\pm 1}^{\dagger }$,

\begin{eqnarray}
H_{\pm 1} &=&\nu E_{0}+\omega _{\pm 1}X_{0}+g_{0}\left( E^{\dag }+E\right)
X_{0}\pm \frac{g_{1}^{2}}{\omega _{\pm 1}+\nu }\left( E^{\dag
2}+E^{2}\right) X_{0}  \label{H11d} \\
&&+\frac{g_{1}}{2}\left( 1+\frac{1}{\Delta _{\pm 1}}\right) \left(
EX_{+}+E^{\dag }X_{-}\right) -g_{0}\frac{g_{1}}{\omega _{\pm 1}+\nu }%
(X_{+}+X_{-})  \notag \\
&&-g_{0}\frac{g_{1}}{\omega _{\pm 1}+\nu }\left( E^{\dag
2}X_{+}+E^{2}X_{-}\right) +\frac{g_{1}}{2}\left( \frac{1}{\Delta _{\pm 1}}%
-1\right) \left( E^{\dag 3}X_{+}+E^{3}X_{-}\right) ,  \notag
\end{eqnarray}%
where $\Delta _{\pm 1}=\sqrt{1\pm 4g_{1}^{2}/(\omega +\nu )^{2}}$, and $%
\omega _{\pm 1}=(\omega +\nu )\Delta _{\pm 1}$.

The elimination of the CR term $\sim X_{+}+X_{-}$ only produces corrections
to the terms already present in (\ref{H11d}), and thus can be neglected,
since we are interested\ only in the principal order of the effective
interaction constants. On the contrary, the elimination of the CR term $\sim
E^{\dag 2}X_{+}+h.c.$ leads to the appearance of $\sim E^{3}X_{0}+h.c.$, $%
\sim E^{\dag 4}X_{+}+h.c.$ and $\sim E^{\dag 5}X_{+}+h.c.$, and in addition
to the modification of the coefficient of $\sim E^{\dag 3}X_{+}+h.c.$. Such
an elimination procedure of CR terms $\sim $ $f_{\pm k}\left( E^{\dag
k}X_{+}+h.c.\right) $, $k=1,2,\ldots $ can be systematically carried out by
applying the transformations
\begin{equation}
V_{\pm k}=e^{A_{\pm k}},\quad A_{\pm k}=\varepsilon _{\pm k}\left( E^{\dag
k}X_{+}-E^{k}X_{-}\right) ,
\end{equation}%
with
\begin{equation}
T\left( 2\varepsilon _{\pm k}\right) =\frac{2f_{\pm k}}{\omega +k\nu }.
\notag
\end{equation}%
An important observation should be made here about the order of $f_{\pm k}$
and $\varepsilon _{\pm k}$
\begin{equation*}
f_{\pm k}\sim g_{0,1}\varepsilon ^{(k-1)},\quad \varepsilon _{\pm k}\sim
\varepsilon ^{(k)},
\end{equation*}%
where $\varepsilon ^{(k)}$, is a homogeneous polynomial of order $k$ on some
small parameters $\varepsilon _{j}\sim g_{0,1}/l.c(\omega ,\nu )\ll 1$,
being $l.c.(\omega ,\nu )$ a linear combination of $\omega $ and $\nu $,
\begin{equation}
\varepsilon ^{(k)}=\sum_{j_{1}\ldots ,j_{s}}c_{j_{1},\ldots
,j_{s}}\varepsilon _{j_{1}}^{n_{j_{1}}}\dots \varepsilon
_{j_{s}}^{n_{j_{s}}},  \label{epsilonk}
\end{equation}%
where $n_{j_{1}}+\dots +n_{j_{s}}=k$, and $c_{j_{1},\ldots ,j_{s}}$ are real
numbers.

Oncethe principal orders of CR terms $\sim $ $\left( E^{\dag
k}X_{+}+h.c.\right) $ are removed we arrive at the following form
\begin{eqnarray*}
H_{\pm 2} &=&\tilde{\omega}X_{0}+\nu E_{0}+g_{1}\left( EX_{+}+E^{\dag
}X_{-}\right)  \\
&&+\sum_{k=1}^{\infty }h_{\pm k}\left( E^{\dag k}+E^{k}\right) X_{0},
\end{eqnarray*}%
where $\tilde{\omega}$ is the system's modified frequency. The couplings $%
h_{\pm k}$ are obtained from the following recurrence relations
\begin{eqnarray*}
h_{\pm 1} &=&g_{0},\quad f_{\pm 1}=g_{1}, \\
h_{\pm k} &=&\pm 2g_{1}\frac{f_{\pm (k-1)}}{\omega +(k-1)\nu }\sim
g_{1}\varepsilon ^{(k-1)},
\end{eqnarray*}%
for $k=2,\ldots $, and
\begin{eqnarray}
f_{\pm (2k+1)} &=&\mp \frac{1}{4g_{1}}\left( h_{\pm (k+1)}\right) ^{2}
\notag \\
&&-\sum_{m=1}^{k}h_{\pm m}\frac{f_{\pm (2k+1-m)}}{\omega +\left(
2k+1-m\right) \nu }\sim g_{1}\varepsilon ^{(2k)},  \notag \\
f_{\pm 2k} &=&-\sum_{m=1}^{k}h_{\pm m}\frac{f_{\pm (2k-m)}}{\omega +\left(
2k-m\right) \nu }\sim g_{1}\varepsilon ^{(2k-1)},  \notag
\end{eqnarray}%
for $k=1,\ldots $. The CR terms of the form $\sim E^{k}X_{0}+h.c$ commute
with each other and can be removed altogether with the transformation%
\begin{equation}
U_{\pm }=\exp \left[ \pm \sum_{k=1}^{\infty }\delta _{\pm k}\left( E^{\dag
k}-E^{k}\right) X_{0}\right] ,  \notag
\end{equation}%
where
\begin{equation*}
\delta _{\pm k}=\frac{h_{\pm k}}{k\nu }\sim \varepsilon ^{(k)},
\end{equation*}%
obtaining the expansion
\begin{equation*}
H_{\pm 3}\approx \tilde{\omega}X_{0}+\nu E_{0}+g_{1}\left( EX_{+}e^{\pm
\sum_{k=1}\delta _{\pm k}\left( E^{\dag k}-E^{k}\right) }+h.c.\right) .
\end{equation*}%
The term $EX_{+}e^{\pm \sum_{k=1}\delta _{\pm k}E^{\dag k}}+h.c$ does not
contribute to the principal order of the effective coupling constants and
can be neglected. \ Then, using the standard expansion
\begin{equation}
\exp \left[ \sum_{k=1}^{\infty }\frac{a_{\pm k}E^{k}}{k!}\right]
=\sum_{k=0}^{\infty }\frac{\mathrm{B}_{k}\left( a_{\pm 1},a_{\pm 2},\ldots
\right) }{k!}E^{k},
\end{equation}%
where $\mathrm{B}_{k}\left( a_{\pm 1},a_{\pm 2},\ldots \right) $ are
complete Bell polynomials \cite{combinatoria}, we finally obtain the
required resonance expansion, which contains only the resonant terms i.e.,
terms that become time-independent under appropriate relations between the
frequencies $\omega $ and $\nu $,

\begin{equation}
H_{ RE\pm}^{F}\approx \tilde{\omega}X_{0}+\nu E_{0}+g_{1}\sum_{k=0}^{\infty
}\epsilon _{\pm k}\left( E^{k+1}X_{+}+E^{k+1}X_{-}\right) ,  \label{Hre1d}
\end{equation}%
where $\epsilon _{\pm 0}=1$,
\begin{equation}
\epsilon _{\pm k}=\frac{\mathrm{B}_{k}\left( a_{\pm 1},a_{\pm 2},\ldots
\right) }{k!},\qquad a_{\pm k}=-k!\delta _{\pm k}  \notag
\end{equation}%
for $n=1,\ldots $.

The resonance expansion (\ref{Hre1d}) contains all possible effective
resonant transitions (resonances) that take place only at $\tilde{\omega}%
\approx (k+1)\nu $. It is noticeable that in the vicinity of every resonance
$\sim \left( E^{k+1}X_{+}+h.c.\right) $ the effect of all the other
resonances can be neglected. In order to see this we remove all the terms
that are non-resonant at $\tilde{\omega}\approx (k+1)\nu $ by applying the
transformation
\begin{equation}
W_{\pm m}=e^{B_{\pm m}},\quad B_{\pm m}=\tilde{\varepsilon}_{\pm m}\left(
E^{m}X_{+}-E^{\dag m}X_{-}\right) ,  \notag
\end{equation}%
\textbf{\ }for $m=1,\ldots $ and $m\neq k+1$, where
\begin{equation}
T\left( 2\tilde{\varepsilon}_{\pm m}\right) =\frac{2g_{1}\epsilon _{\pm m-1}%
}{\omega -m\nu },  \notag
\end{equation}%
to the expansion (\ref{Hre1d}). This results in the following effective
Hamiltonian describing the resonant transition $\tilde{\omega}\approx
(k+1)\nu $ implicitly present in the Hamiltonian (\ref{H1dop0}),
\begin{eqnarray}
H_{eff\pm }^{F} &\approx &\tilde{\omega}X_{0}+\nu E_{0}+\frac{g_{1}}{2}\left[
\frac{1}{\tilde{\Delta}_{\pm m}}+1\right] \sum_{m-1\neq k=0}^{\infty
}\epsilon _{\pm k}\left( E^{k+1}X_{+}+E^{\dag k+1}X_{-}\right)
\label{HeffA1} \\
&&+\frac{g_{1}}{2}\left[ \frac{1}{\tilde{\Delta}_{\pm m}}-1\right]
\sum_{m-1\neq k=0}^{\infty }\epsilon _{\pm k}\left( E^{2m}E^{\dag
k+1}X_{+}+E^{\dag 2m}E^{k+1}X_{-}\right)   \label{HeffA2} \\
&&+\frac{g_{1}^{2}\epsilon _{\pm m}}{(\omega -m\nu )\tilde{\Delta}_{\pm m}}%
\sum_{m-1\neq k=0}^{\infty }\epsilon _{\pm k}\left( E^{m}E^{\dag
k+1}+E^{\dag m}E^{k+1}\right) X_{0},  \label{HeffA3}
\end{eqnarray}%
where $\tilde{\Delta}_{\pm m}=\sqrt{1\pm 4g_{1}^{2}\epsilon _{\pm
m-1}^{2}/(\omega -m\nu )^{2}}$. It can observed that the modified frequency $%
\tilde{\omega}$ is changed, but the principal order of the coupling constant
corresponding to the resonant term $\sim \left( E^{k+1}X_{+}+h.c.\right) $
in (\ref{HeffA1}) remains the same as in (\ref{Hre1d}). All of the other
terms (\ref{HeffA2})-(\ref{HeffA3}) generate contributions of smaller order.

\section{Two coupled systems with modulated frequency}

\label{B}

In this Appendix we obtain the resonance expansion corresponding to the
Hamiltonian (\ref{H2sistgent}),
\begin{equation}
H^{F}=\omega _{0}X_{0}+\frac{1}{2}\omega _{0}\gamma \left( E^{\dag
}+E\right) X_{0}+\omega _{1}Y_{0}+\nu E_{0}+g\left( X_{+}+X_{-}\right)
\left( Y_{+}+Y_{-}\right) .
\end{equation}%
First, by applying the transformation (\ref{TransA}), with $\epsilon =\omega
_{0}\gamma /\nu \lesssim 1$ we obtain
\begin{eqnarray}
VH^{F}V^{\dagger } &=&\omega _{0}X_{0}+\omega _{1}Y_{0}+\nu
E_{0}+gJ_{0}(\epsilon )\left( X_{+}+X_{-}\right) \left( Y_{+}+Y_{-}\right)
\notag \\
&&+g\sum_{k=1}^{\infty }J_{k}(\epsilon )\left[ X_{+}\left( E^{\dag
k}+(-1)^{k}E^{k}\right) +h.c.\right] \left( Y_{+}+h.c.\right) .
\label{Hf2sys}
\end{eqnarray}%
Now we consequtively apply the set of transformations
\begin{equation}
V_{1k}=\exp \left[ J_{k}(\epsilon )\varepsilon _{k}\left( E^{\dag
k}X_{+}Y_{+}-h.c.\right) \right] ,\;k=0,1,\ldots ,  \notag
\end{equation}%
where $\varepsilon _{k}=g/(\omega _{0}+\omega _{1}+k\nu )$, to the
Hamiltonain (\ref{Hf2sys}) in order to remove CR terms $J_{k}(\epsilon
)\left( E^{\dag k}X_{+}Y_{+}+h.c.\right) $ in the weak coupling limit, $g\ll
\omega _{0,1}$. The transformed Hamiltonain contains, in addition to the
resonant terms, CR contributions of the form: $\sim \varepsilon
^{(1)}E^{\dag k}X_{+}^{2}+h.c.$, $\sim \varepsilon ^{(1)}E^{\dag
k}Y_{+}^{2}+h.c.$, $\sim \varepsilon ^{(2)}E^{\dag k}X_{+}^{3}Y_{+}+h.c.$, $%
\sim \varepsilon ^{(2)}E^{\dag k}X_{+}Y_{+}^{3}+h.c.$ y $\sim \varepsilon
^{(1)}E^{\dag k}\nabla _{x,y}\Phi (X_{0},Y_{0})+h.c.$, where%
\begin{eqnarray}
\Phi (X_{0},Y_{0}) &=&\phi _{x}(X_{0})\phi _{y}(Y_{0}),  \label{Phi} \\
\nabla _{mx,ny}f(X_{0},Y_{0}) &=&f(X_{0},Y_{0})-f(X_{0}+m,Y_{0}+n).
\label{nabla2}
\end{eqnarray}%
After eliminating all those CR terms we eventually arrive at a resonance
expansion that contains a diagonal contribution $K(X_{0},Y_{0})$ as an
important ingredient. The operator $K(X_{0},Y_{0})$depends non-linearly  on $%
X_{0}$ and $Y_{0}$, except for the case when both $X$ and $Y$ systems are
described by $h(1)$ algebra and can be interpreted as an intensity-dependent
frequency shift. Up to third order on small parameters $\varepsilon _{k}\ll 1
$, it has the form
\begin{eqnarray}
K(X_{0},Y_{0}) &\approx &\varepsilon ^{(1)}(\epsilon )\nabla _{x,y}\Phi
(X_{0},Y_{0})  \label{P2gen} \\
&&+\varepsilon ^{(3)}(\epsilon )\nabla _{x,y}\left[ \Phi (X_{0},Y_{0})\nabla
_{x,y}^{2}\Phi (X_{0}-1,Y_{0}-1)\right]   \label{P31gen} \\
&&+\varepsilon ^{(3)}(\epsilon )\left[ \left( \nabla _{y}\phi
_{y}(Y_{0})\right) ^{2}\nabla _{2x}\left( \phi _{x}(X_{0})\phi
_{x}(X_{0}-1)\right) \right]   \label{P32gen} \\
&&+\varepsilon ^{(3)}(\epsilon )\left[ \left( \nabla _{x}\phi
_{x}(X_{0})\right) ^{2}\nabla _{2y}\left( \phi _{y}(Y_{0})\phi
_{y}(Y_{0}-1)\right) \right] ,  \label{P33gen}
\end{eqnarray}%
where
\begin{equation}
\varepsilon ^{(m)}(\epsilon )=\sum_{l_{1},\ldots ,l_{s}}c_{l_{1},\ldots
,l_{s}}J_{l_{1}}^{n_{l_{1}}}(\epsilon )\dots J_{l_{s}}^{n_{l_{s}}}(\epsilon
)J_{l_{s+1}}(\epsilon )\varepsilon _{l_{1}}^{n_{l_{1}}}\dots \varepsilon
_{l_{s}}^{n_{l_{s}}},  \label{epsilonkJ}
\end{equation}%
$n_{l_{1}}+\dots n_{l_{s}}=m$, are some homogeneous polynomials of the
Bessel functions.

The intensity dependent frequency shift (\ref{P2gen})-(\ref{P33gen})
automatically suppresses higher-order transitions leading to the excitation
of X and Y systems. Since we consider only $h(1)$, $su(2)$ and $su(1,1)$
algebras, the maximum degree of $\Phi (X_{0},Y_{0})$ on $X_{0}$ and $Y_{0}$
is four. Thus, the resonance expansion that includes only possible efficient
transitions takes the form%
\begin{equation*}
H_{RE}^{F}\approx \omega _{0}X_{0}+\omega _{1}Y_{0}+\nu
E_{0}+gK(X_{0},Y_{0})+H_{int},
\end{equation*}%
where the effective interacion Hamiltonian $H_{int}$ has the following
structure%
\begin{equation*}
H_{int}=\sum_{k=1}\sum_{m=0}^{3}\sum_{n}\varepsilon _{nk}^{(m)}H_{nk}^{(m)},
\end{equation*}%
and $H_{nk}^{(m)}$ are given in the following Tables
\begin{equation*}
\begin{array}{|c|c|c|}
\hline
H_{nk}^{(0)} & H_{nk}^{(1)} & H_{nk}^{(2)} \\ \hline
J_{0}(\epsilon )\left( X_{+}Y_{-}+h.c.\right)  & E^{k}X_{+}^{2}\nabla
_{y}\phi _{y}(Y_{0})+h.c. & E^{k}X_{+}^{3}Y_{+}\nabla _{y}^{2}\phi
_{y}(Y_{0})+h.c. \\
J_{k}(\epsilon )\left( E^{\dag k}X_{+}Y_{-}+h.c.\right)  &
E^{k}Y_{+}^{2}\nabla _{x}\phi _{x}(X_{0})+h.c. & E^{k}X_{+}Y_{+}^{3}\nabla
_{x}^{2}\phi _{x}(X_{0})+h.c. \\
J_{k}(\epsilon )\left( E^{k}X_{+}Y_{-}+h.c.\right)  &  & E^{k}X_{+}^{3}Y_{-}%
\nabla _{y}^{2}\phi _{y}(Y_{0})+h.c. \\
J_{k}(\epsilon )\left( E^{k}X_{+}Y_{+}+h.c.\right)  &  & E^{k}Y_{+}^{3}X_{-}%
\nabla _{x}^{2}\phi _{x}(X_{0})+h.c. \\
&  & E^{\dag k}X_{+}^{3}Y_{-}\nabla _{y}^{2}\phi _{y}(Y_{0})+h.c. \\
&  & E^{\dag k}Y_{+}^{3}X_{-}\nabla _{x}^{2}\phi _{x}(X_{0})+h.c \\ \hline
\end{array}%
\end{equation*}%
\begin{equation*}
\begin{array}{|c|}
\hline
H_{nk}^{(3)} \\ \hline
E^{k}X_{+}^{2}Y_{+}^{2}\nabla _{x,y}^{3}\Phi (X_{0},Y_{0})+h.c. \\
E^{k}X_{+}^{2}Y_{+}^{2}\nabla _{-2y}\nabla _{y}\phi _{y}(X_{0})\nabla
_{2x}\nabla _{x}\phi _{x}(X_{0})+h.c. \\
E^{k}X_{+}^{2}Y_{+}^{2}\nabla _{2y}\nabla _{y}\phi _{y}(X_{0})\nabla
_{-2x}\nabla _{x}\phi _{x}(X_{0})+h.c. \\
E^{k}X_{+}^{2}Y_{+}^{2}\nabla _{-2y}\nabla _{y}\phi _{y}(X_{0})\nabla
_{2x}\nabla _{x}\phi _{x}(X_{0})+h.c. \\
E^{k}X_{+}^{2}Y_{-}^{2}\nabla _{2y}\nabla _{y}\phi _{y}(X_{0})\nabla
_{2x}\nabla _{x}\phi _{x}(X_{0})+h.c. \\
E^{k}X_{+}^{2}Y_{-}^{2}\nabla _{2y}\nabla _{y}\phi _{y}(X_{0})\nabla
_{2x}\nabla _{x}\phi _{x}(X_{0})+h.c. \\
E^{k}X_{+}^{2}Y_{+}^{2}\nabla _{2y}\nabla _{y}\phi _{y}(X_{0})\nabla
_{-2x}\nabla _{x}\phi _{x}(X_{0})+h.c. \\
E^{k}X_{+}^{2}Y_{+}^{2}\nabla _{2y}\nabla _{y}\phi _{y}(X_{0})\nabla
_{-2x}\nabla _{x}\phi _{x}(X_{0})+h.c. \\
E^{k}X_{+}^{2}Y_{-}^{2}\nabla _{2y}\nabla _{y}\phi _{y}(X_{0})\nabla
_{2x}\nabla _{x}\phi _{x}(X_{0})+h.c. \\
E^{k}X_{+}^{2}Y_{-}^{2}\nabla _{2y}\nabla _{y}\phi _{y}(X_{0})\nabla
_{2x}\nabla _{x}\phi _{x}(X_{0})+h.c. \\
E^{k}X_{+}^{2}Y_{-}^{2}\nabla _{2y}\nabla _{y}\phi _{y}(X_{0})\nabla
_{2x}\nabla _{x}\phi _{x}(X_{0})+h.c. \\
E^{k}X_{+}^{2}Y_{-}^{2}\nabla _{2y}\nabla _{y}\phi _{y}(X_{0})\nabla
_{2x}\nabla _{x}\phi _{x}(X_{0})+h.c. \\
E^{\dag k}X_{+}^{2}Y_{-}^{2}\nabla _{2y}\nabla _{y}\phi _{y}(X_{0})\nabla
_{2x}\nabla _{x}\phi _{x}(X_{0})+h.c. \\
E^{k}X_{+}^{3}\nabla _{x,-y}\left[ \phi _{x}(X_{0})\nabla _{y}^{2}\phi
_{y}(Y_{0}+1)\right] +h.c. \\
E^{k}Y_{+}^{3}\nabla _{-x,y}\left[ \phi _{y}(Y_{0})\nabla _{x}^{2}\phi
_{x}(X_{0}+1)\right] +h.c. \\ \hline
\end{array}%
\end{equation*}%
Higher orders of the interaction Hamiltonians contain higher discrete
derivatives of the structural functions $\phi _{x}(X_{0})$ and $\phi
_{y}(Y_{0})$. This in particular, allows to determine all possible
resonances when both $X$ and $Y$ systems are described by $h(1)$ algebras.

\subsection{Non-symmetric excitation of an atomic system in a vacuum}

\label{b1}

Applying an elimination procedure similar to that described in Appendix A to
the Hamiltonian (\ref{DMda}) we arrive at the following resonance expansion
up to the second order on $\epsilon $

\begin{eqnarray}
H_{RE}^{F} &\approx &\sum_{i=1}^{2}\left( \omega _{i}+g_{i}\varepsilon
_{1i}\right) s_{zi}+\omega _{c}a^{\dag }a+\nu E_{0}  \notag \\
&&+2g_{1}\sum_{i=1}^{2}\left( \omega _{i}+g_{i}\varepsilon _{1i}\right)
a^{\dag }as_{zi}-g_{2}\varepsilon _{11}\left( s_{-1}s_{+2}+h.c.\right)
\notag \\
&&+\sum_{i=1}^{2}g_{i}\left[ \left( 1+\epsilon \left( E^{\dag }-E\right) +%
\frac{\epsilon ^{2}}{2}\left( E^{\dag 2}+E^{2}\right) \right) as_{+i}+h.c.%
\right]  \notag \\
&&-\sum_{i=1}^{2}g_{i}\left[ \left( E-\frac{\epsilon }{2}E^{2}\right)
a^{\dag }s_{+i}+h.c.\right] -\sum_{i=1}^{2}g_{i}\varepsilon _{1i}\varepsilon
_{2i}\left( a^{3}s_{+i}+h.c.\right)  \notag \\
&&+2\sum_{i=1}^{2}g_{i}\epsilon \varepsilon _{1i}\left( E^{\dag
}a^{2}+h.c.\right) s_{zi}  \notag \\
&&+\epsilon \left( g_{1}\varepsilon _{12}-g_{2}\varepsilon _{11}\right)
\left( E^{\dag }s_{-1}s_{+2}-Es_{-1}s_{+2}+h.c.\right)  \notag \\
&&+\epsilon \left( g_{1}\varepsilon _{12}+g_{2}\varepsilon _{11}\right)
\left( Es_{+1}s_{+2}+h.c.\right) ,  \label{2atom}
\end{eqnarray}
where $\epsilon =g_{0}/\nu $, $\varepsilon _{1i}=g_{i}/(\omega _{c}+\omega
_{i})$, and $\varepsilon _{2i}=g_{i}/\omega _{c}$.

\end{document}